\documentclass{aastex62}
\usepackage{amsmath,amssymb,CJK,soul}

\received{--}
\revised{--}
\accepted{--}
\shorttitle{Detection of Small NEAs with ZTF}
\shortauthors{Ye et al.}
\begin{document}
\begin{CJK*}{UTF8}{gbsn}

\title{Towards Efficient Detection of Small Near-Earth Asteroids Using the Zwicky Transient Facility (ZTF)}

\correspondingauthor{Quanzhi Ye}
\email{qye@caltech.edu}

\author[0000-0002-4838-7676]{Quanzhi Ye (叶泉志)}
\affiliation{Division of Physics, Mathematics and Astronomy, California Institute of Technology, Pasadena, CA 91125}
\affiliation{IPAC, California Institute of Technology, 1200 E. California Blvd, Pasadena, CA 91125}

\author{Frank J. Masci}
\affiliation{IPAC, California Institute of Technology, 1200 E. California Blvd, Pasadena, CA 91125}

\author[0000-0001-7737-6784]{Hsing~Wen~Lin \begin{CJK*}{UTF8}{bkai}(林省文)\end{CJK*}}
\affiliation{Department of Physics, University of Michigan, Ann Arbor, MI 48109}
\affiliation{Institute of Astronomy, National Central University, 32001, Taiwan}

\author[0000-0002-4950-6323]{Bryce Bolin}
\altaffiliation{B612 Asteroid Institute and DIRAC Institute Postdoctoral Fellow}
\affiliation{Department of Astronomy, University of Washington, Seattle, WA 98195}
\affiliation{B612 Asteroid Institute, 20 Sunnyside Ave, Suite 427, Mill Valley, CA 94941}

\author[0000-0003-1656-4540]{Chan-Kao Chang \begin{CJK*}{UTF8}{bkai}(章展誥)\end{CJK*}}
\affiliation{Institute of Astronomy, National Central University, 32001, Taiwan}

\author[0000-0001-5060-8733]{Dmitry A. Duev}
\affiliation{Division of Physics, Mathematics and Astronomy, California Institute of Technology, Pasadena, CA 91125}

\author{George Helou}
\affiliation{IPAC, California Institute of Technology, 1200 E. California Blvd, Pasadena, CA 91125}

\author{Wing-Huen~Ip \begin{CJK*}{UTF8}{bkai}(葉永烜)\end{CJK*}}
\affiliation{Institute of Astronomy, National Central University, 32001, Taiwan}
\affiliation{Space Science Institute, Macau University of Science and Technology, Macau}

\author[0000-0001-6295-2881]{David L. Kaplan}
\affiliation{Department of Physics, University of Wisconsin-Milwaukee, Milwaukee, WI 53211}

\author[0000-0003-0457-2519]{Emily Kramer}
\affiliation{Jet Propulsion Laboratory, California Institute of Technology, Pasadena, CA 91109}

\author[0000-0003-2242-0244]{Ashish Mahabal}
\affiliation{Division of Physics, Mathematics and Astronomy, California Institute of Technology, Pasadena, CA 91125}
\affiliation{Center for Data Driven Discovery, California Institute of Technology, Pasadena, CA 91125}

\author[0000-0001-8771-7554]{Chow-Choong Ngeow}
\affiliation{Institute of Astronomy, National Central University, 32001, Taiwan}

\author{Avery J. Nielsen}
\affiliation{Division of Physics, Mathematics and Astronomy, California Institute of Technology, Pasadena, CA 91125}

\author{Thomas A. Prince}
\affiliation{Division of Physics, Mathematics and Astronomy, California Institute of Technology, Pasadena, CA 91125}

\author{Hanjie Tan (谭瀚杰)}
\affiliation{Department of Physics, National Central University, 32001, Taiwan}

\author{Ting-Shuo Yeh \begin{CJK*}{UTF8}{bkai}(葉庭碩)\end{CJK*}}
\affiliation{Institute of Astronomy, National Central University, 32001, Taiwan}

\author[0000-0001-8018-5348]{Eric C. Bellm}
\affiliation{DIRAC Institute, Department of Astronomy, University of Washington, 3910 15th Avenue NE, Seattle, WA 98195}

\author{Richard Dekany}
\affiliation{Caltech Optical Observatories, California Institute of Technology, Pasadena, CA 91125}

\author{Matteo Giomi}
\affiliation{Institute of Physics, Humboldt-Universit\"at zu Berlin, Newtonstr. 15, 12489 Berlin, Germany}

\author[0000-0002-3168-0139]{Matthew J. Graham}
\affiliation{Division of Physics, Mathematics and Astronomy, California Institute of Technology, Pasadena, CA 91125}

\author[0000-0001-5390-8563]{Shrinivas R. Kulkarni}
\affiliation{Division of Physics, Mathematics and Astronomy, California Institute of Technology, Pasadena, CA 91125}

\author[0000-0002-6540-1484]{Thomas Kupfer}
\affiliation{Kavli Institute for Theoretical Physics, University of California, Santa Barbara, CA 93106}

\author{Russ R. Laher}
\affiliation{IPAC, California Institute of Technology, 1200 E. California Blvd, Pasadena, CA 91125}

\author{Ben Rusholme}
\affiliation{IPAC, California Institute of Technology, 1200 E. California Blvd, Pasadena, CA 91125}

\author{David L. Shupe}
\affiliation{IPAC, California Institute of Technology, 1200 E. California Blvd, Pasadena, CA 91125}
             
\author{Charlotte Ward}
\affiliation{Department of Astronomy, University of Maryland, College Park, MD 20742}



\begin{abstract}
We describe ZStreak, a semi-real-time pipeline specialized in detecting small, fast-moving near-Earth asteroids (NEAs) that is currently operating on the data from the newly-commissioned Zwicky Transient Facility (ZTF) survey. Based on a prototype originally developed by \citet{Waszczak2017} for the Palomar Transient Factory (PTF), the predecessor of ZTF, ZStreak features an improved machine-learning model that can cope with the $10\times$ data rate increment between PTF and ZTF. Since its first discovery on 2018 February 5 (2018 CL), ZTF/ZStreak has discovered $45$ confirmed new NEAs over a total of 232 observable nights until 2018 December 31. Most of the discoveries are small NEAs, with diameters less than $\sim100$~m. By analyzing the discovery circumstances, we find that objects having the first to last detection time interval under 2~hr are at risk of being lost. We will further improve real-time follow-up capabilities, and work on suppressing false positives using deep learning.
\end{abstract}



\section{Introduction}

Small Solar System bodies are remnants of the formation stage of the Solar System. They encompass small natural objects in the Solar System with sizes from $\sim1$~meter to a few hundred kilometers, including near-Earth objects (NEOs), main-belt asteroids, trans-Neptunian objects, and various other smaller groups of asteroids and comets. Studies of small bodies contribute to the understanding of several fundamental questions in planetary science, such as the composition of the proto-planetary disk, the evolutionary history of Solar System, as well as the transportation and distribution of water and organic materials in the Solar System.

NEOs are of particular interest because, aside from a purely scientific perspective, they also pose threats to our civilization and provide opportunities for resource utilization for future space activities. For example, it is now widely recognized that the impact of a 10-km-class asteroid is responsible for the extinction of the dinosaurs \citep[c.f.][]{Schulte2010}. Since the 1990s, a handful of dedicated NEO surveys have dramatically increased our knowledge of the NEO population. According to the statistics released by the International Astronomical Union Minor Planet Center (MPC)\footnote{\url{https://www.minorplanetcenter.net/}}, more than 19,000 NEOs had been discovered by the end of 2018. The majority of known NEOs are asteroids (collectively known as near-Earth asteroids or NEAs), i.e. the rocky and relatively dry component of the remnants from the proto-planetary disk; only $\sim100$ known NEOs are comets.

\begin{figure}
\includegraphics[width=0.5\textwidth]{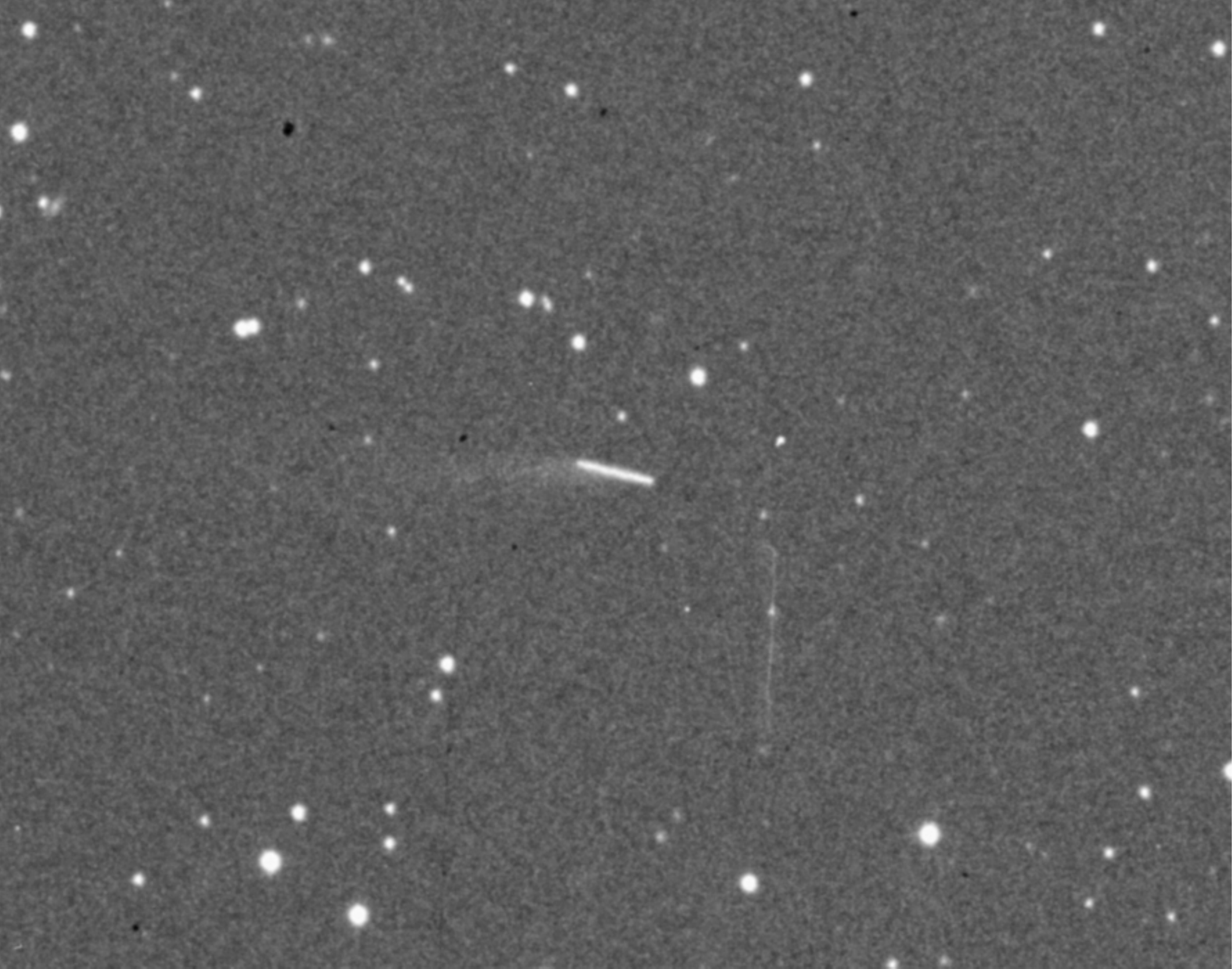}
\caption{A streaked object (center). Shown here is 107P/(4015) Wilson-Harrington, the first NEO discovered by the Palomar 1.2-m Oschin Schmidt telescope (Palomar 48-inch, P48), showing up as a streaked object on its discovery image taken on 1949 November 19 \citep[less than two months after the first light of the telescope, c.f.][]{Bowen1949, Cunningham1950}. Despite appearing as a streak, Wilson-Harrington was actually moving at a rate of only 2$^\circ$/day at that time, and was streaked merely because of the long exposure (12 minutes) of the image.\label{fig:107p}}
\end{figure}

At this point, our knowledge of the distribution of km-sized NEOs is fairy complete \citep{Jedicke2015}, but drops sharply towards smaller sizes. Asteroids that are tens to $\sim100$-m in size are generally not expected to cause global catastrophe, but are still capable to cause severe damage on city or larger scales \citep{Binzel1997}. The most recent and notable example is probably the Chelyabinsk event in 2013, which caused significant property damage to the city of Chelyabinsk in Russia, but was produced by an previously undiscovered asteroid that was only 18-m in diameter \citep{Borovivcka2013, Brown2013}. Telescopic detection of these small asteroids is challenging, since they are either very faint when they are far from the Earth, or they have high apparent motion rates when close and bright enough to be detected (see Figure~\ref{fig:107p} for an example). NEOs that approach the Earth within $\sim15$ lunar distances typically move at a rate of $>10^\circ$/day \citep{Verevs2012}. These ``Fast-Moving Objects'' (FMOs) would trail on typical survey exposures (usually 20--60 seconds) and present a challenge for traditional NEO detection algorithms, which are tuned to detect objects moving slower than a few degrees per day \citep{Jedicke2013}.

Searching for asteroids  by searching for ``streaking'' objects dates back to the late 19th century, as virtually all discoverable asteroids back then would move noticeably during the hour-long exposure using photographic plates. (433) Eros, the first NEA found in 1898, was found as a long streak on a 2-hour exposure \citep{Scholl2002}. The term ``Fast-Moving Object'' more specifically came into use  to refer to streaked NEOs in the 1970s \citep[e.g.][]{Aksnes1971, Helin1976a, Morrison1976}, as the operation of larger telescopes as well as the use of more sensitive photographic films reduced the exposure time. The application of highly sensitive charge-coupled device (CCD) to NEO surveys, pioneered by the Spacewatch survey in the early 1990s \citep{Rabinowitz1991a, Scotti1991}, virtually limits the FMO phenomenon to objects that pass very close to the Earth. While modern NEO surveys are largely automatic and mostly only requires human attention in the quality assurance stage, detection of streaking FMOs remains a challenge and still requires more human intervention. In 2004--2006, Spacewatch had successfully conducted a citizen science project (the Spacewatch FMO Project) that allowed the public to access their survey images and identify possible FMOs, for which the survey observers would initiate follow-up observations to refine the orbit \citep{McMillan2005}.

Using the survey data from the Palomar Transient Factory (PTF), \citet[][hereafter W17]{Waszczak2017} successfully demonstrated a prototype pipeline dedicated to FMO detection. PTF was a synoptic survey operated using the 1.2-m Oschin Schmidt telescope (Palomar 48-inch, P48) and the 7.3~deg$^2$ CFH12K camera at Palomar Observatory from 2009 to 2016 \citep{Law2009, Rau2009}. PTF was succeeded by the Zwicky Transient Facility \citep[ZTF; ][]{Bellm2019} in 2018 which replaced the CFH12K camera with a dedicated 47~deg$^2$ camera. Routine NEO searches using ZTF data have been initiated, with emphasis on FMO detection. Here we present the ongoing effort on implementing the FMO detection capability on ZTF.

\section{Overview of the ZTF Survey}

ZTF is specifically designed for visiting the entire northern visible sky (north of $-30^\circ$ declination for Palomar) every night in one filter.
%GH every three nights in both $g'$ and $r'$ filters, and at a higher cadence in selected regions (Graham et al. in prep). 
The new 576 megapixel camera (Dekany et al. in prep), installed on the 1.2-m Oschin Schmidt telescope, can observe an area of 3760 deg$^2$ in an hour, to a $5\sigma$ median detection limit of $m_{r'}=20.7$ with a 30-second integration. A total of 40\% of the observation time is devoted to public surveys, with the visible northern sky observed in $g'$ and $r'$ every three nights, and the visible Galactic Plane ($|b|<7^\circ$) observed in $g'$ and $r'$ every night. The remaining 60\% time (comprised of 40\% partnership time and 20\% Caltech time) is divided into a number of sub-surveys, with the major one being an hourly-cadence survey of mid-declination fields \citep{Bellm2019}.

Data acquired by the telescope are transmitted to Infrared Processing and Analysis Center (IPAC) at Caltech and are processed in real-time \citep{Masci2019}. IPAC manages  the central ZTF Science Data System (ZSDS), handling the calibration and processing of single epoch images, astrometry and photometry, image co-addition, image differencing, alert generation, and moving objects. The moving object branch is divided into the ZTF Moving Object Detection Engine (ZMODE) that specializes in linking point-source detections of moving objects, and the ZTF Streak pipeline (ZStreak) which focuses on the detection of streaked objects. Although sharing largely the same scientific objectives, ZMODE is executed independently of ZStreak. The technical aspects and some initial results from ZMODE are covered in \citet{Masci2019} and \citet{Graham2019}.

\section{Detection Process}

\begin{figure*}
\includegraphics[width=\textwidth]{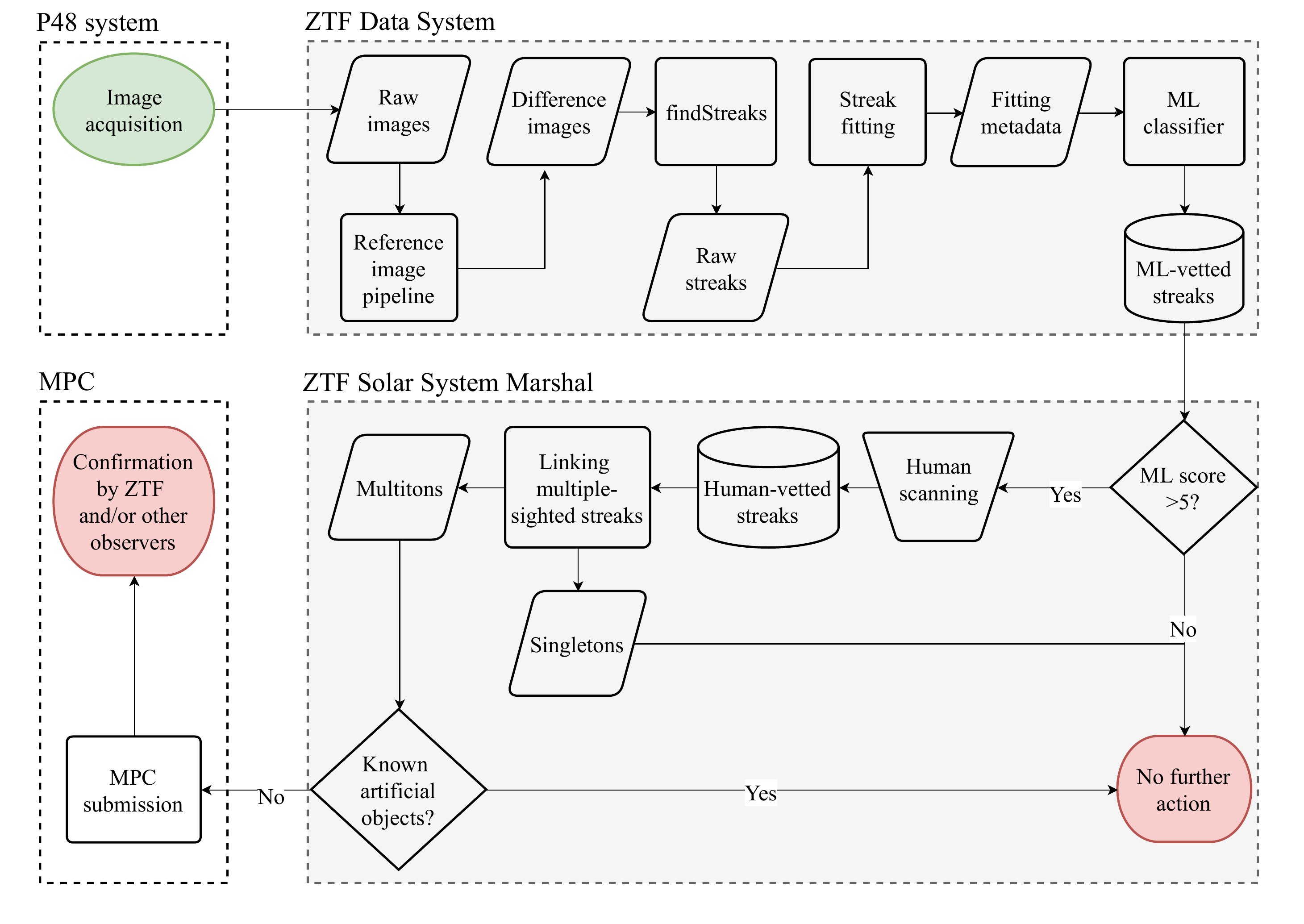}
\caption{Flowchart showing the process of FMO discovery.\label{fig:flow}}
\end{figure*}

The streak detection process is outlined in Figure~\ref{fig:flow} and is further described below:

\subsection{Initial Detection} 

The first step of the detection process is to search for linear image features that resemble the appearance of ``trailed'' sources in difference images. As described in \citet[][\S~3.6]{Masci2019}, the difference images are generated by subtracting ``reference'' images from the single epoch science images. The reference images are ``static'' representations of the sky, constructed from co-adding historical science images acquired earlier in the survey. The use of difference images suppresses the majority of static sources, which would otherwise lead to large numbers of false positives and contaminate the streak detection process.

The search for FMO candidates is first performed by using the {\it findStreaks} software, which is documented in the original W17 paper in the context of PTF. {\it findStreaks} identifies candidates by searching for contiguous bright pixels that exceed a signal-to-noise threshold and whose spatial distribution is approximately linear according to  an estimate of the Pearson correlation coefficient. The software generates a set of metadata describing the morphology of each identified ``raw'' streak, including equatorial coordinates of its midpoint, length, width, positional angle, and integrated flux. Compared to the W17 work, {\it findStreaks} has been optimized for ZTF, primarily to improve the detection efficiency (completeness) for fainter, shorter streaks. The correlation coefficient threshold was reduced to 0.55 (formerly 0.65) and the minimum length reduced to 3 pixels (formerly 9 pixels). The pixel signal-to-noise threshold remains at 1.5$\sigma$. For ZTF, the effective minimum length of candidate streaks reported by {\it findStreaks} (following all internal thresholding) is $\approx5$~pixels (or 5'', which corresponds to a minimum on-sky motion of 4$^\circ$/day for typical ZTF exposures). For PTF, the shortest reported lengths were $\approx10''$ (or 4$^\circ$/day for typical PTF exposures). Ensuring we can detect shorter streaks with ZTF is important since the nominal integration time of the individual exposures is half (30 sec) of what it was for PTF. In practice, we find that candidate streaks from ZTF images shorter than $\approx5''$ turn out to be spurious (unreliable) detections. We compensate for this by setting a lower cutoff of 7~pixels (5$^\circ$/day) when selecting candidates for inclusion in the training set for the machine-learned classifier. This is further described in \S~\ref{sect:ml}.

The original W17 work used the metrics from {\it findStreaks} as direct input to their machine-learned classifier, with astrometric and flux information therefrom included in MPC submissions. Here we improve the estimation and reporting process by introducing an additional step to further characterize the candidate streaks.

\subsection{Streak Point-Spread Function (PSF) Fitting Code}

Photons from FMOs, like those from other astronomical sources, are subjected to the point-spread function (PSF) effect of the telescope system. This has two implications: (1) PSF-related metrics can be used as an effective discriminator against certain image artifacts such as cosmic rays, internal reflections of the optical system, and sensor artifacts; and (2) construction of the FMO's PSF profile can allow better astrometry to be derived, as demonstrated in previous work \citep[e.g.][]{Rabinowitz1991a, Kouprianov2008, Verevs2012}.

To derive metrics that describe the PSF behavior of a streak, we apply the trail fitting technique described in \citet{Verevs2012} and \citet{Lin2015}: the base function is essentially a finite-length streak convolved with a Gaussian PSF of width $\sigma$, which can be written as

\begin{equation}
\label{eq:psf}
f(x',y') = b + \frac{F}{L} \frac{1}{2\sigma\sqrt{2\pi}} \exp{\left(-\frac{y'^2}{2\sigma^2}\right)} \
\left[ \mathrm{erf} \left( \frac{x'+L/2}{\sigma\sqrt{2}} \right) - \mathrm{erf} \left( \frac{x'-L/2}{\sigma\sqrt{2}} \right) \right]
\end{equation}

\noindent where $b$ is the background level, $F$ is the total integrated flux of the streak, $L$ is the length of the streak, and

\begin{displaymath}
\mathrm{erf} (z) = \frac{2}{\pi} \int_0^z \exp{\left(-t^2\right)}\mathrm{d} t
\end{displaymath}

\noindent is the Gaussian error function, and

\begin{align*}
x' = (x-x_0) \cos{\theta} + (y-y_0) \sin{\theta} \\
y' = (x-x_0) \sin{\theta} + (y-y_0) \cos{\theta}
\end{align*}

\noindent where $x_0$, $y_0$ defines the centroid of the streak, and $\theta$ defines the angle between the motion of the streak and the $x$ axis. Note that a finite shutter opening and closing time of 0.43 second has not been included in this PSF.

\begin{figure*}
\includegraphics[width=\textwidth]{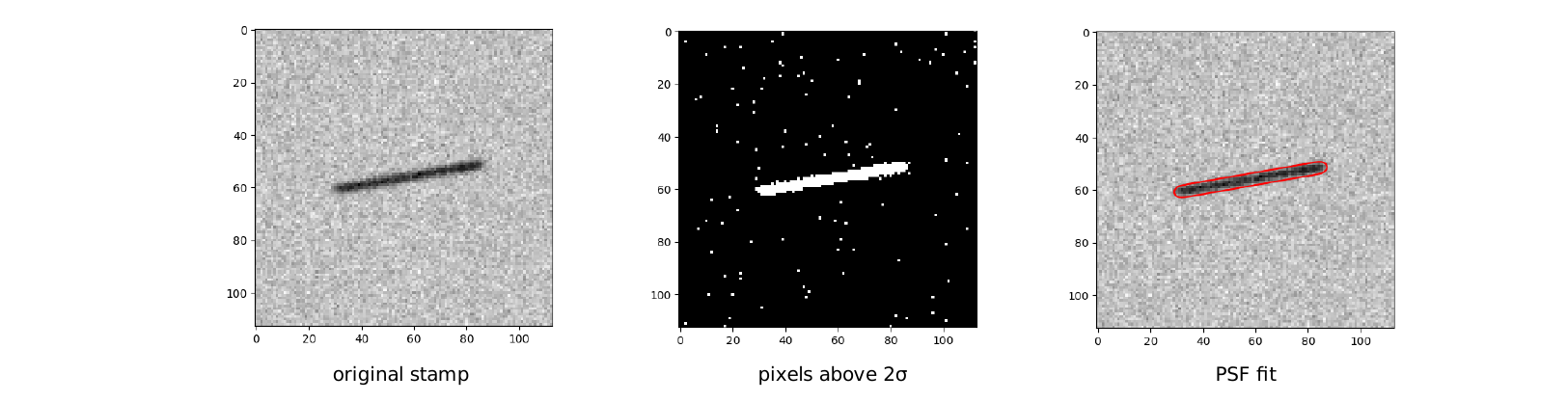}
\caption{Streak PSF fitting: original cutout (left), pixels above $2\sigma$ level (middle), and the fitting result indicated by the contour (right). The axes are in pixels; 1 pixel $\approx1''$. \label{fig:psf-fitting}}
\end{figure*}

Metrics calculated by {\it findStreaks} are used as inputs to the streak fitting code. The code attempts to fit the image feature, defined by pixels above $2\sigma$ level, using the Levenberg-Marquardt technique (Figure~\ref{fig:psf-fitting}). We start with the initial position suggested by {\it findStreaks}, and perform a walk within the $3\times3$ grid around the initial positions if the fit fails, until the fit is successful. Experiments show that streaks that cannot be fitted after multiple attempts do not exhibit any PSF-like behavior. These streaks are therefore rejected. For the ones that can be fitted, the code will compute the $F$, $L$, $x_0$, $y_0$, $\sigma$, $\theta$ and $b$ for each of these streaks, and append these variables to the metadata previously computed by {\it findStreaks}.

After PSF-fitting filtering, one typical ZTF exposure will yield several hundreds streaks. A full night of ZTF observations will yield an order of $10^5$ streaks. The majority of these are long satellite tracks, bright star diffraction spikes, and various image artifacts. This is $\sim10\times$ reduction/improvement over PTF, which would produce a similar number of raw streaks \citep{Waszczak2017} but with $10\times$ smaller field-of-view. Still, screening $10^5$ candidates by eye is a challenging task. Therefore, we follow W17's approach and use machine learning (ML) to cope with this challenge.

\subsection{Machine-learned classifier} \label{sect:ml}

Similar to W17, we use a scikit-learn-based Random Forest classifier \citep{Pedregosa2012} to set up our ML model, but have redefined the feature list as well as the method to construct the training set \citep[see also][]{Mahabal2019}.

W17 defined 15 features derived from {\it findStreaks}, and used a hybrid of 1,441 real FMOs and synthetically-generated streaks, as well as $\sim20,000$ bogus streaks (i.e., raw streaks that are not real FMOs) as their training set. We define 15 features derived from the PSF fitting code described above (Table~\ref{tbl:ml-feature}) that mostly do not overlap with the feature list defined by W17. We also do not use real FMOs in our training set, as they represent the brighter asteroid population that may introduce bias into the classifier. The relatively rare occurrence of real FMOs ($\lesssim 10$ real FMOs per night, including different sighting of the same object) also lead to the possible issue of over-fitting. Instead, we generate $\sim50,000$ synthetic streaks using Equation~\ref{eq:psf} and inject them into random ZTF differenced images, and extract their morphological metrics as we would for the raw streaks. We then randomly select $\sim50,000$ ZTF raw streaks from ZTF difference images and label them as bogus streaks. The contamination of real FMOs in the bogus set is expected to be minimal, again due to the rare occurrence of real FMOs.

\begin{table*}
\begin{center}
\caption{Features derived by the PSF fitting code and used for the ML training. \label{tbl:ml-feature}}
\begin{tabular}{lll}
\tableline
Feature name & Description & Variable in Eq.~\ref{eq:psf} \\
\tableline
flux & Integrated aperture flux of the streak & $F$ \\
bg & Local background level & $b$ \\
length & Length of the streak & $L$ \\
sigma & Width of the streak  & $\sigma$ \\
lengtherr & Error of the length of the streak  & $L$ \\
sigmaerr & Error of the width of the streak & $\sigma$ \\
paerr & Error of the positional angle of the streak & $\theta$ \\
bgerr & Error of the local background level & $b$ \\
fitmagerr & Error of the fit magnitude & \\
apsnr & SNR of the flux within the aperture & \\
apmagerr & Error of the aperture magnitude & \\
dmag & Difference between the aperture magnitude and the PSF-fitted magnitude & \\
dmagerr & Error of dmag & \\
chi2 & $\chi2$ of the fit & \\
numfit & Number of attempts for a converged fit & \\
\tableline
\end{tabular}
\end{center}
\end{table*}

The ML model is then trained using the $\sim50,000$-sample ``real'' set (though here it is really a ``synthetic'' set) and the $\sim50,000$-sample ``bogus'' set. We set the number of trees to 1,000 after experimenting with performance versus resource cost. Other ML-related parameters (such as depth of a tree, number of samples per leaf, and number of leaf nodes) are left unconstrained, as they do not have a noticeable effect on the performance. The synthetics are derived assuming a flat distribution of motion direction, a flat distribution of angular motion rate between 5--50$^\circ$/day, and an apparent brightness distribution following $N\propto F^{-0.65}$ \citep[derived from the power-law distribution of sub-km NEAs derived by][]{Mainzer2011} within a range of $V=15$ to $V=20$, where $N$ is the cumulative number of synthetics to be generated and $F$ is the flux level. The brightness distribution is derived following the consensus on the NEO size distribution \citep[c.f.][]{Jedicke2015}. Note that the ``true'' distribution is more complicated: the brightness distribution and the motion rate is dependent to each other, what we construct here is merely a zero-order approximation to the problem.

Model training was carried out more frequently during the commissioning months of ZTF (first half of 2018), due to construction and re-construction of reference images (which may slightly affect the quality of resulting differenced images), as well as engineering modifications to the ZTF system. As a result, the numbers in each of the sets vary a bit. As of this writing, the most recent model was trained and passed to the Data System pipeline in mid-June of 2018.

The ML model computes a score that varies from 0 to 1, with 0 corresponding to a streak  least likely to be real while 1 corresponds to the most likely. Currently, we adopt a threshold of 0.05, which (as will be shown below) is $98\%$ complete in detecting real FMOs presented in the raw streak sample. The detection completeness up to this stage will be discussed in greater detail in \S~\ref{sec:process:completeness}.

\subsection{Human scanning and streak-linking}

\begin{figure*}
\includegraphics[width=\textwidth]{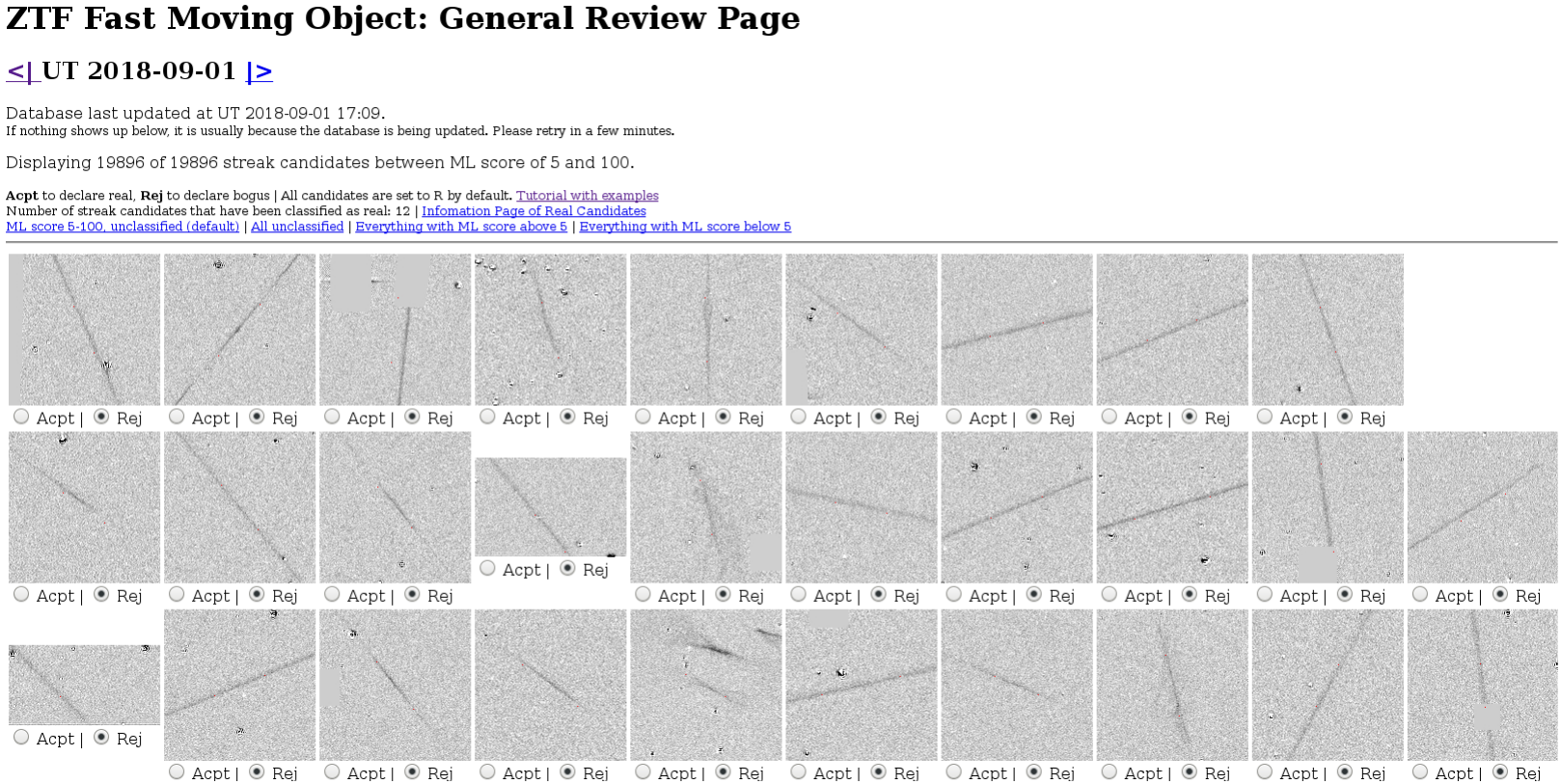}
\caption{ZStreak general scanning page.\label{fig:scanning-page}}
\end{figure*}

Streaks with scores higher than the ML score threshold are posted on a webpage (Figure~\ref{fig:scanning-page}) for human scanning. Typical lag time from image acquisition to posting of ML-vetted streaks is 20--40~minutes. A typical night would see $\sim10^4$ streaks accumulated on this page, which is $\sim10\%$ of the raw streaks initially detected by {\it findStreaks}.

Presently, scanning is done twice in a night: a mid-night (Pacific time) scan, usually carried out by scanners residing in a convenient timezone (e.g. Europe, Asia); and a morning scan, usually carried out by scanners in the U.S. The duties of the scanner are to identify and save objects that are potentially interesting for further inspection. For experienced scanners, it usually takes up to 20~minutes to review a half night's worth of data ($\sim5$--$10\times10^4$).  The plan is to gradually distribute the scanning effort over the Pacific night, in order to reduce elapsed time between exposure and identification of potentially interesting objects.

\begin{figure*}
\includegraphics[width=\textwidth]{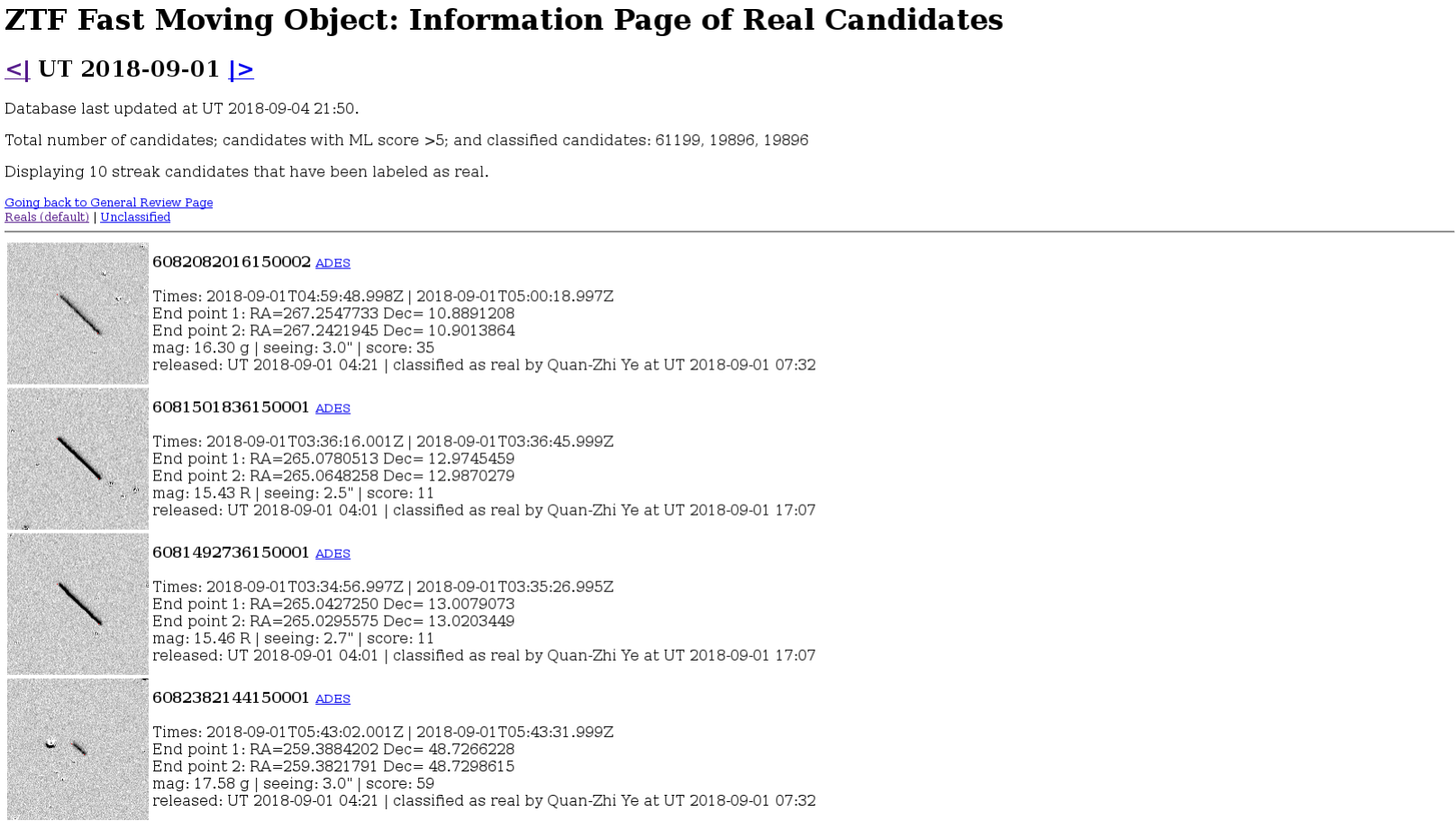}
\caption{The webpage that displays additional information for streaks that are potentially interesting.\label{fig:shepherd-page}}
\end{figure*}

Once the scanning is done, a separate page displays additional information for objects that are potentially interesting (see Figure~\ref{fig:shepherd-page}). This information  allows operators to identify streaks that are potentially different sightings of the same object. Operators  compare the position, direction and length of the streaks, and identify streaks that have similar direction, length, and region of appearance. These are streaks that are likely linked to each other. Plausible linkages are further examined using Find\_Orb, an open-source orbit-computing software\footnote{\url{https://www.projectpluto.com/find_orb.htm}.}, to see if they have converging orbit solutions.

The idea of streak-linking is similar to the linking of tracklets in the conventional ZMODE-like point-source moving object detection \citep[c.f.][]{Denneau2013, Verevs2017}; however, it should be noted that each streak detection contains time, position \textit{and} velocity information, as opposed to point-source object detection, which only contains time and position. The variable-in-question in a streak detection is the direction of the motion, which has two solutions and requires a secondary detection to fully determine. Therefore, while streak-linking only requires a minimum of two detections to operate, ZMODE-like system requires a minimum of three.

The converged orbital solution of two or more streaks is called a ``track''. Each track is first  checked against the database of known artificial objects and asteroids, using the sat\_id tool\footnote{\url{https://www.projectpluto.com/sat_id.htm}.} and the NEO Checker tool\footnote{\url{https://www.minorplanetcenter.net/cgi-bin/checkneo.cgi}.}. The MPC does not have the responsibility of collecting observations of artificial objects, and therefore observations of these objects are not  submitted. Observations of both known and unknown asteroids are immediately submitted to the MPC. The MPC discourages labeling known asteroids\footnote{\url{https://www.minorplanetcenter.net/iau/info/Astrometry.html\#id}, retrieved 2018 September 1.}, therefore known asteroids are submitted by their temporary identifiers which are of the form ZTFXXXX (where X can be 0-9, A-Z and a-z).

\subsection{Discovery Announcement and Follow-up}

Unlike  the procedure outlined in W17 of triggering Target-of-opportunity (ToO) sessions for streaks with a single detection, we generally only consider objects from which a track can be generated. This is primarily due to the benefit we can reap from ZTF's large field-of-view, which is sufficient to detect most FMOs more than once under typical survey cadence. For plausible FMOs that are sighted only once, human operators will search for secondary detections in the images of the same region taken near the detection epoch. For most of these cases, no secondary detection can be found even though nearby images exist, suggesting that these are in fact image artifacts or satellite flashes. On rare occasions, the images containing secondary sightings did not go through the reference image pipeline and therefore did not generate streak products. This is usually due to the lack of pre-existing reference images in the field \citep{Masci2019}, and we expect that such a situation will become less common as ZTF accumulates data over the sky. Sometimes ZStreak can also miss the secondary sighting even though it is in the image, due to a variety of reasons (e.g., interference from nearby bright star or satellite tracks, being close to the edges, or simply being too faint). In these cases, the secondary sightings are manually measured and are added to the track.

Observations are submitted to the MPC in the new Astrometry Data Exchange Standard (ADES) format\footnote{\url{https://www.minorplanetcenter.net/iau/info/ADES.html}.}. The ADES format implements data fields that are not represented in the traditional MPC1992 format, such as astrometric and photometric uncertainty, both of which can be derived from the PSF fitting code and other ZTF data products. Incorporating astrometric uncertainty is particularly important for FMO observations, since the along-streak uncertainty is typically larger than cross-streak uncertainty and is expected to have an impact on the quality of the resulting computed orbit.

Upon receipt of the observations, the MPC performs its own check against known asteroids and artificial objects. If both turn out to be negative, the object is posted on the Near-Earth Object Confirmation Page (NEOCP) to facilitate follow-up observations from observers worldwide. The ZTF partnership typically uses the 1-m telescope at Lulin Observatory \citep[][]{Ye2009b, Lin2018} located on the other side of the Pacific for follow-up purposes. In principle, the P48 system would only conduct self follow-up if the future positional uncertainty of an object using two or more streaks would be too large for telescopes with typical field-of-view to recover (see discussions in \S~\ref{results:disc}).

\section{Preliminary Results}

\subsection{NEAs} \label{results:disc}

ZStreak discovered its first FMO (2018 CL) on 2018 Febuary 5, the first night of its test operation \citep{Ye2018}. As of 2018 December 31, ZStreak has discovered a total of 45 FMOs over a total of 232 observable nights. (A ``discovery'' is defined as a FMO that has been assigned a provisional designation by the MPC and is first reported by ZTF.) Most (36 out of 45) were found in the second half of the year, as the first half of the year saw extensive modifications and updates as part of the ZTF commissioning phase. The operation in the second half of the year lost a month (October) due to an instrumental issue. Meanwhile, software is continuously being improved. Therefore we believe that we are yet to reach the full power of ZStreak.

\begin{figure}
\includegraphics[width=\textwidth]{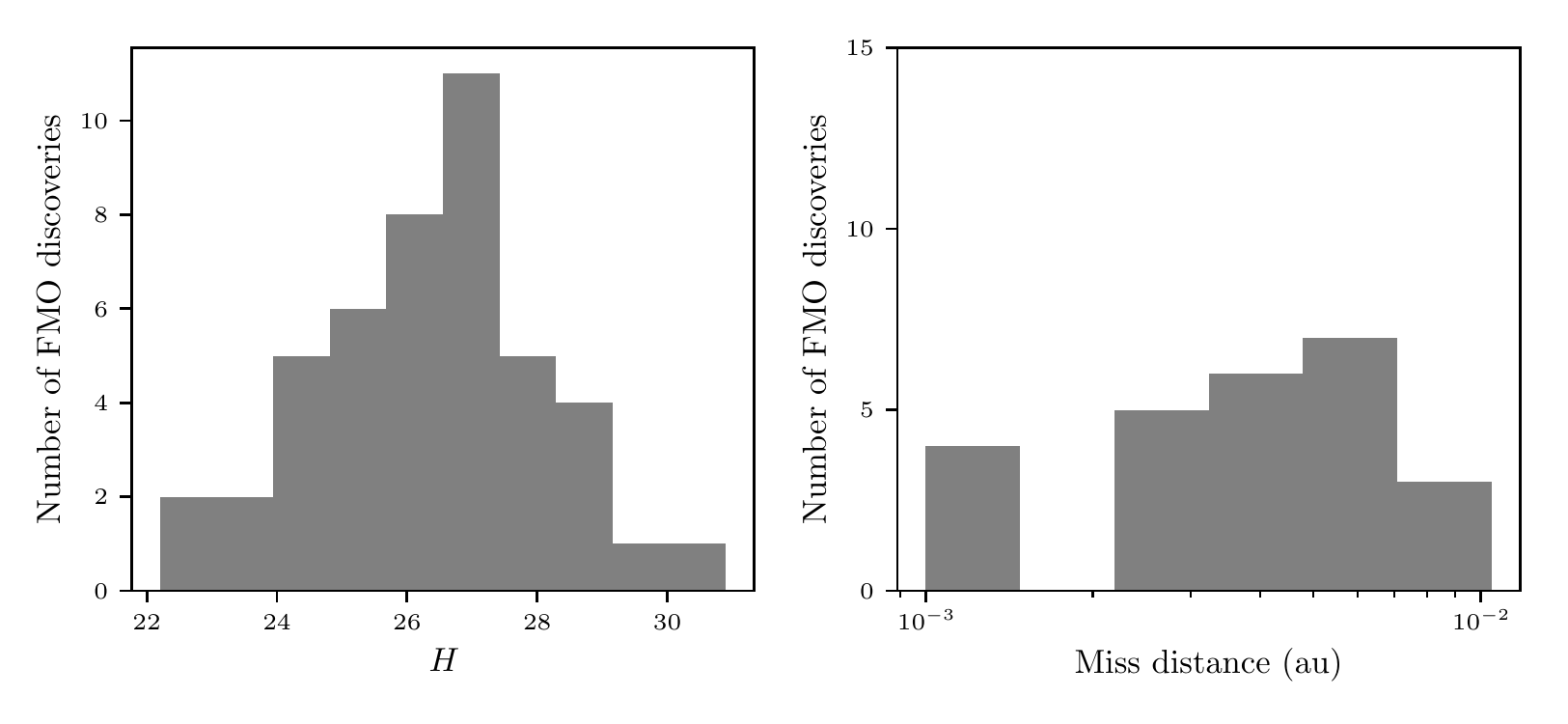}
\caption{Distribution of absolute magnitudes ($H$) and miss distances of confirmed ZStreak discoveries. \label{fig:h-dist}}
\end{figure}

To understand the NEA population that ZStreak is probing, we analyze the miss distances to the Earth and the absolute magnitudes of the discovered FMOs, with the result shown as Figure~\ref{fig:h-dist}. Unsurprisingly, ZStreak is mostly finding small NEAs that would soon pass or have just passed very close ($\lesssim0.02$ au, or 10~LD) to the Earth. The mean absolute magnitude of the ZStreak discoveries is $\bar{H}=26.2\pm1.7$ (corresponding to a diameter of $\sim30$~m assuming an albedo of 0.1), which is higher than those of other NEO surveys \citep[$23.3\pm2.6$, c.f.][]{Veres2018}. This highlights the fact that ZTF/ZStreak preferentially finds smaller NEAs due to its capability in detecting streaks.

Some objects posted on NEOCP did not receive adequate follow-up observations and were subsequently declared as ``lost'' NEOs. As of 2018 December 31, a total of 53 ZStreak objects posted on NEOCP were lost due to insufficient follow-ups. This indicates a loss rate of $\sim50\%$, comparable to earlier numbers derived by W17, but is much higher than average for other NEO surveys \citep[$11\%$, c.f.][]{Veres2018}.

\begin{figure}
\includegraphics[width=0.5\textwidth]{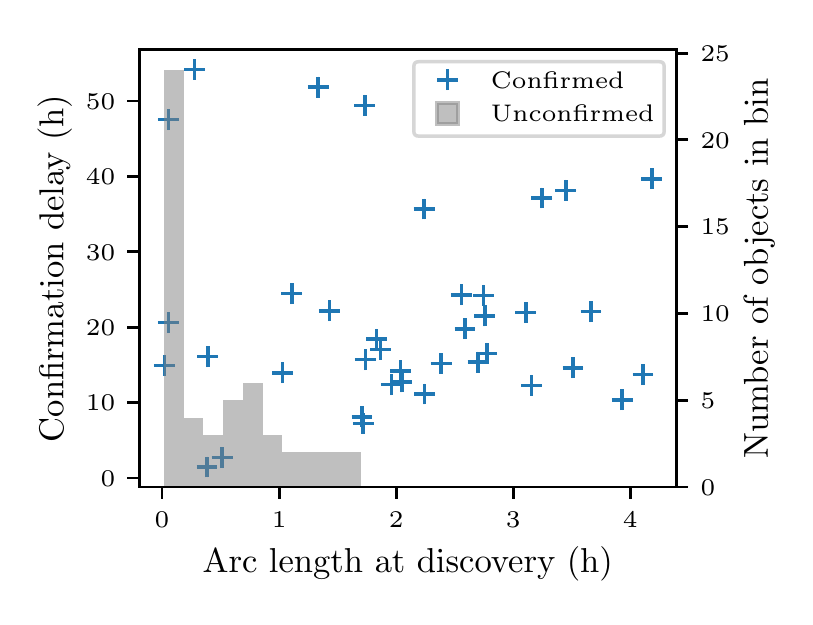}
\caption{Relation between delay of independent confirmation (for the confirmed ZTF NEAs) and the arc length at initial ZTF detection (for both confirmed and unconfirmed ZTF NEAs). Objects that are considered ``confirmed'' by the MPC but did not receive any independent confirmation are not shown here.\label{fig:arc-confdelay}}
\end{figure}

\begin{figure}
\includegraphics[width=0.5\textwidth]{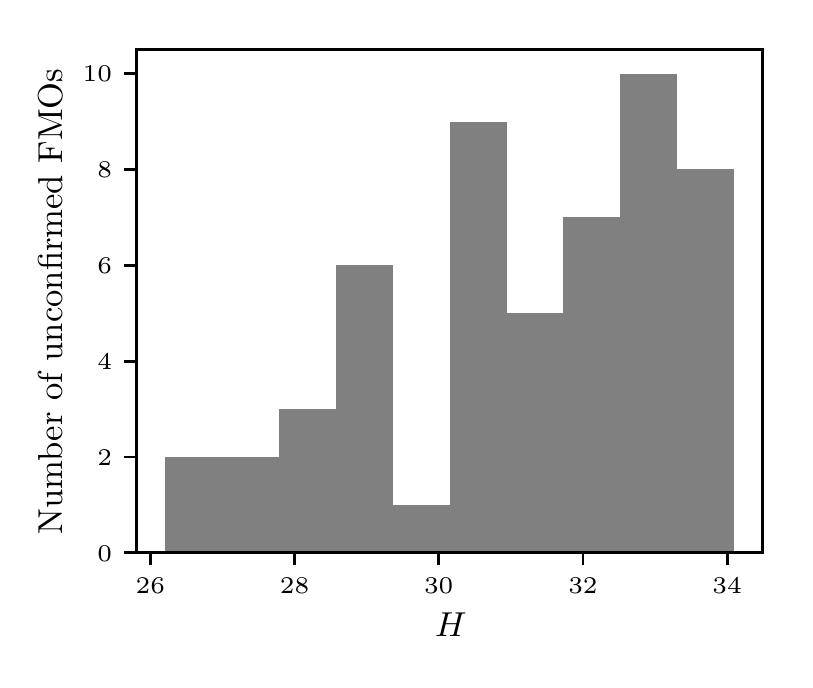}
\caption{Absolute magnitude ($H$) of the unconfirmed FMOs found by ZStreak.\label{fig:unconfirmed-H}}
\end{figure}

To determine the cause of lost FMOs, we analyze the length of the orbital arc of the discovery data for both confirmed and unconfirmed objects, as well as the delay of the first confirmation for confirmed objects. As shown in Figure~\ref{fig:arc-confdelay}, we find that most confirmed discoveries have discovery arc length of over two hours, while \textit{all} unconfirmed discoveries have discovery arc length $<2$~hr. (The discovery arc length is the time span between the first and the last detection of the discovery.) This suggests that the population of confirmed NEAs are biased towards NEAs that have longer discovery arc lengths. This could be helped by triggering additional target-of-opportunity observations 2 hours after the initial detection. Additional analysis of the $H$ of the unconfirmed objects shows $\bar{H}=31.1$\footnote{The $H$ of an unconfirmed object is simply derived from the orbit solution with minimum $O-C$ difference. We note that the orbit of an unconfirmed object is often poorly constrained due to the limited length of the arc.}, indicating that these are very small, meter-sized objects (Figure~\ref{fig:unconfirmed-H}).

\subsection{Completeness} \label{sec:process:completeness}

The completeness of a survey is limited by a variety of factors, such as sky coverage, detection efficiency of the source-finding software, sufficient visits to establish a tracklet, and so on. Modern-day NEO surveys heavily rely on automatic software, therefore their completeness is largely limited by software capabilities. Compared to point-source-based method, FMO detection has its own unique niche and challenge: it requires fewer images to build up tracklets and tracks, but is more vulnerable to the contamination from other sources and artifacts since a FMO would spread across multiple pixels.

To quantify the completeness of ZStreak, we compile a list of known FMOs that are imaged by ZTF from 2018 February 5 to 2018 May 11, and then re-process the ZTF images containing these FMOs, using the current version of ZStreak. We then examine the raw streaks produced by the re-processing to estimate the likelihood of detection at each sighting of these FMOs.

\begin{figure*}
\includegraphics[width=\textwidth]{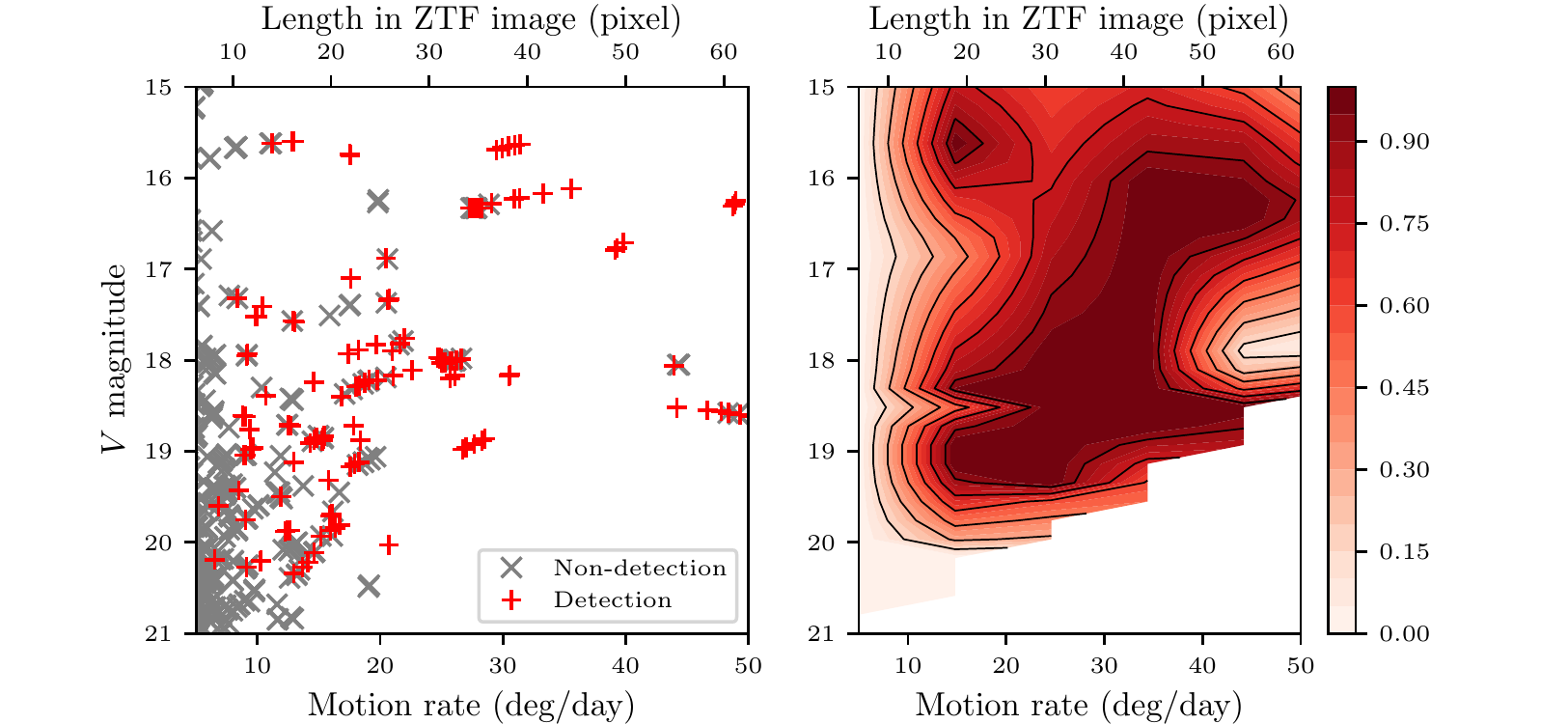}
\caption{Completeness of ZStreak detection using known FMOs observed by ZTF from 2018 February 5 to May 11. Left panel shows a distribution of individual objects, and right panel shows linearly-interpolated density map. \label{fig:completeness}}
\end{figure*}

Figure~\ref{fig:completeness} shows the completeness of ZStreak for individual streak detection derived from 265 positive detections of known (i.e. MPC-cataloged) FMOs. We find that ZStreak works well with FMOs faster than $\sim10^\circ$/day, which are approximately 15 pixels long at typical ZTF exposures (30-second). The overall completeness of {\it findStreaks} \textit{and} PSF-fitting filtering is $\sim72\%$ for FMOs that are between a $V$ magnitude of 15 and 20 and are $>12^\circ$/day, which is noticeably higher than the $45\%$ completeness achieved by W17 in the same magnitude range. Most of the improvement comes from the detection of fainter FMOs: W17 reported a detection limit of $V\sim18.5$, while our detection limit reaches $V\sim19.8$. The improvement of completeness is primarily due to a cleaner background, as well as the fine tuning of {\it findStreaks}.

\begin{figure*}
\includegraphics[width=\textwidth]{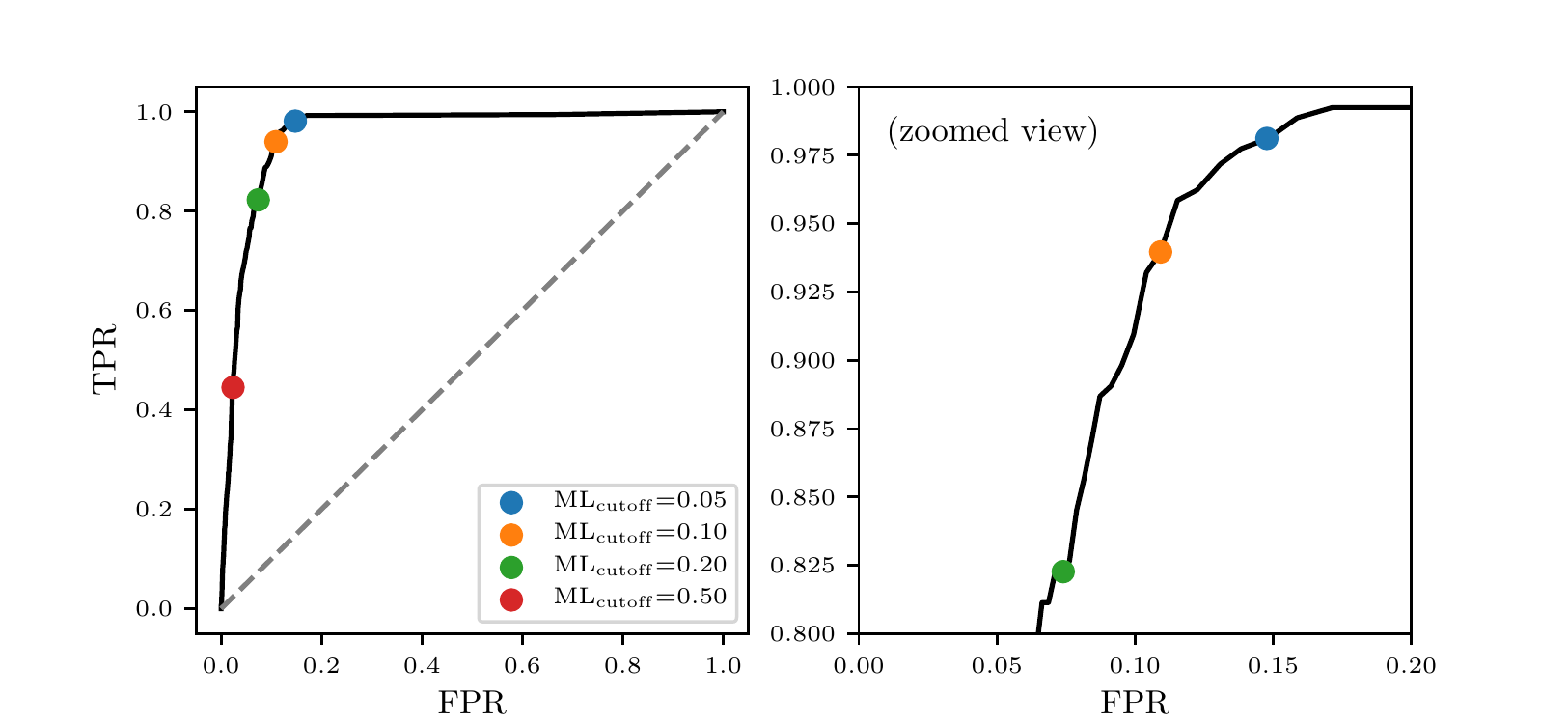}
\caption{Performance of the ML classifier as revealed by the receiver operating characteristic (ROC) curve. Shown here is the full ROC curve (left) as well as a zoomed plot of it (right). Also shown is the benchmarks for different cutoff score for the ML classifier. \label{fig:roc}}
\end{figure*}

Another potential bottleneck that can limit the completeness is the stage of ML classification. To assess the performance of the ML, we compute the true positive rate (TPR; i.e., fraction of real streaks correctly classified as such) using the positive detections of known FMOs collected above, as well as the false positive rate (FPR; i.e., fraction of bogus streaks incorrectly classified as reals) from a random query of the bogus streaks in the ZStreak database. The resulting receiver operating characteristic (ROC) plot is shown as Figure~\ref{fig:roc}. We find that a ML score cutoff at the level of 0.05  yields a TPR of 0.98 and a FPR of 0.15, which we believe is a good balance for efficiency. Experience from survey operations (since June 2018) shows that the TPR is indeed very close to 1 while the FPR varies from 0.1 to about 0.3. 

The numbers presented above are the completeness of single-sighting detections; the ``true'' completeness of ZTF/ZStreak as an NEO discovery engine will need to take the linkage of multiple-sighted streaks into account. This cannot be calculated easily, since ZTF employs multiple survey strategies. The problem is further complicated by the fact that the human operator occasionally conducts visual searches for the second sightings of single detections that are likely undetected by ZStreak.

Here we provide some rough estimates of the true completeness of the system. For public surveys, which visit each field twice each night (with one in $g'$ and the other in $r'$), the likelihood of successfully detecting a link-able FMO is $\eta=P^2=(72\%)^2=52\%$ (where $P=72\%$ is the likelihood of positive detection of a FMO in the image, as derived above). For sub-surveys that visit the same field $n$ times a night, the likelihood follows 

\begin{displaymath}
\eta=\sum_{k=2}^{n} {n\choose k} P^k (1-P)^{n-k}
\end{displaymath}

For $n\geq3$, $\eta\geq81\%$. If we take the fact that ZTF spends 40\% time on public surveys and na\"{i}vely assumes that ZTF spends the other 60\% on hourly-cadence survey (i.e. $\sim6$ visits per field, per night), a very rough estimate on the global completeness of ZTF/ZStreak is $\sim80\%$.

\subsection{Artificial Objects}

ZStreak's specialty makes it capable of detecting artificial satellites and debris in high Earth orbits (HEOs). Objects in HEOs orbit the Earth beyond the geosynchronous orbit, or 36,000~km (0.09~LD) above Earth's surface, therefore can behave like Earth-approaching asteroids, especially near their apogees. Positions of these objects can be calculated from their respective Two-Line Element (TLE) files, which are made available by the North American Aerospace Defense Command (NORAD) and space enthusiasts.

On the other hand, satellites that exhibit non-Keplerian motion, as well as smaller objects that do not appear in the public catalog, can be confused with new Earth-approaching asteroids. Small objects are particularly challenging since their dynamics are dominated by radiation force and are highly complicated \citep[e.g.][]{Schildknecht2008}. It is often not possible to distinguish artificial objects from NEAs based on the discovery data. These objects will be submitted to the MPC for further confirmation, as follow-up observations are needed to constrain the orbit and to see whether the orbit is geocentric (a strong indicator that the object is artificial).

As of 2018 December 31, a total of 30 ZStreak detections posted on NEOCP are removed for having geocentric orbits (i.e. likely artifical in origin). This suggests that approximately $30/(30+45)=40\%$ of confirmed ZStreak submissions are geocentric objects. In parallel, we find that about half of the unconfirmed candidates noted in \S~\ref{results:disc} have orbit solutions compatible with geocentric objects. It should be note that objects on geocentric orbits are preferentially more difficult to recover, as they are closer to the Earth than typical NEAs. Therefore, we estimate that $\sim50\%$ of ``real'' ZStreak detections are likely artificial in origin.

\begin{figure}
\includegraphics[width=0.5\textwidth]{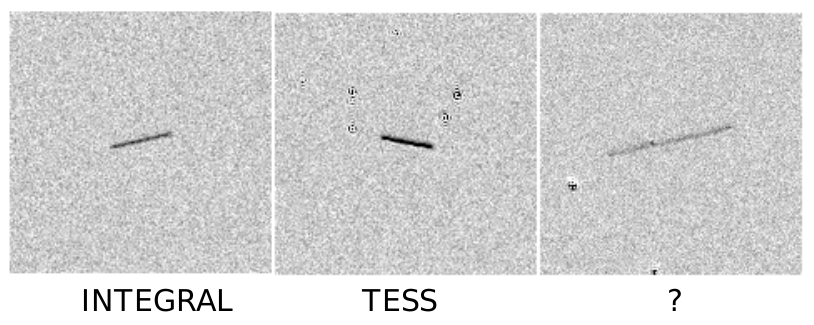}
\caption{Discovery stamps of three artificial objects, from left to right: INTEGRAL, TESS, and an unidentified object.\label{fig:artsat}}
\end{figure}

Figure~\ref{fig:artsat} shows a few examples of HEOs detected by ZStreak. It can be seen that their appearance is indistinguishable from real NEAs. Though some ``new'' objects have been identified to be boosters or fragments from previous HEO missions, many new detections remain unidentified. We note that all unidentified objects have $H>31$, which seems to imply the completeness of the public HEO catalog is somewhere near $H=31$ (about 1--2 meter in diameter, assuming an albedo of 0.5). To that end, a notable discovery is ZTF00hg. At $H=35.4$ (approximately 15~cm in diameter, assuming an albedo of 0.5), this object is probably the smallest object ever detected by a NEO survey.

\section{Conclusions and Future Work}

We have presented the design, operation and preliminary results from ZStreak, a novel pipeline dedicated to real-time detection of fast-moving objects (FMOs) with the new Zwicky Transient Facility (ZTF) survey. We showed that ZStreak achieved a completeness of $72\%$ for FMOs with a $V$ magnitude between 15 and 20, moving at an apparent rate of $>12^\circ$/day. At the time of the writing, ZStreak has operated in semi-real-time since ZTF commissioning (2018 February), and has detected about 100~FMOs in the process, with about half of them assigned to ZTF by the MPC.

It was also found that close to $50\%$ of ZTF-found new FMOs could not be confirmed. We found that objects with initial detection arc length shorter than 2~hours are at risk of being lost. We concluded that an efficient FMO survey with ZTF-like setup will need to have an internal turnaround time (i.e. from telescopic detection to MPC submission) less than $\sim2$~hours to maximize the recovery rate.

Apart from further streamlining and automating the operation, efforts have been focused on reducing the high false positive rate. Among other options \citep[e.g.][]{Lieu2018, Nir2018}, we have been exploring neural networks as an alternative approach to screen raw streaks (Duev et al. in prep). Preliminary testing has shown a $20-25\times$ reduction on the false positive rate. It is hoped that a smaller false positive rate would make it easier to perform real-time scanning, that could help facilitate near-real-time follow-up to prevent discoveries from getting lost.

Looking into the future, the experience of ZTF/ZStreak will provide a useful knowledge base for next generation sky surveys such as the Large Synoptic Survey Telescope (LSST). The smaller pixels, better seeing, and fainter limiting magnitude of LSST will push the detection of trailed objects into a new regime. FMO detection will also be useful for next generation space-based surveys such as {\it NEOCam}, {\it WFIRST} and {\it Euclid} \citep{Bauer2018, Carry2018, Mainzer2016}. Even though {\it WFIRST} and {\it Euclid} are not designed for asteroid detection, being able to extract asteroid observation from the survey data will compliment existing ground- and space-based NEO surveys, as shown by previous works that make use of general-purpose sky surveys \citep[e.g.][]{Ivezic2001, Masiero2011}.

\acknowledgments

We thank an anonymous referee for helpful comments. We extend our thanks to many professional and amateur astronomers around the world for their valuable and timely follow-up observations which are critical for FMO discoveries. We thank Bill Gray in particular for his find\_orb and sat\_id software, which has been very useful to our work. We also thank Eric Christensen, Davide Farnocchia and Gareth Williams for beneficial discussions.

Q.-Z. Ye acknowledges support by the GROWTH (Global Relay of Observatories Watching Transients Happen) project funded by the National Science Foundation PIRE (Partnership in International Research and Education) program under Grant No 1545949. Q.-Z. Ye is grateful for the Spacewatch FMO Project which sparked his interest in asteroids.

C.-C.~Ngeow thanks the funding from Ministry of Science and Technology (Taiwan) under grant 104-2923-M-008-004-MY5.

Based on observations obtained with the Samuel Oschin Telescope 48-inch Telescope at the Palomar Observatory as part of the Zwicky Transient Facility project. ZTF is supported by the National Science Foundation under Grant No. AST-1440341 and a collaboration including Caltech, IPAC, the Weizmann Institute for Science, the Oskar Klein Center at Stockholm University, the University of Maryland, the University of Washington, Deutsches Elektronen-Synchrotron and Humboldt University, Los Alamos National Laboratories, the TANGO Consortium of Taiwan, the University of Wisconsin at Milwaukee, and Lawrence Berkeley National Laboratories. Operations are conducted by COO, IPAC, and UW. 

This research has made use of data and/or services provided by the International Astronomical Union's Minor Planet Center. The National Geographic Society - Palomar Observatory Sky Atlas (POSS-I) image of 107P/(4015) Wilson-Harrington presented in this paper were obtained from the Mikulski Archive for Space Telescopes (MAST). STScI is operated by the Association of Universities for Research in Astronomy, Inc., under NASA contract NAS5-26555. POSS-I was made by the California Institute of Technology with grants from the National Geographic Society.

\facilities{PO:1.2m}
\software{Scikit-learn \citep{Pedregosa2012}}

\bibliographystyle{aasjournal}
\bibliography{ms}



\end{CJK*}
\end{document}